\documentclass[journal]{IEEEtran}
\usepackage{graphicx, amssymb}
\usepackage{subcaption}
\usepackage[backend=biber,style=ieee]{biblatex}
\begin{document}
%###################################
\title{ConFormer: A Novel Collection of Deep \\ Learning Models to Assist Cardiologists \\ in the Assessment of Cardiac Function}
\author{Ethan~Thomas, Salman Aslam% <-this % stops a space
\thanks{Ethan Thomas is a researcher at Ransom Everglades School}
\thanks{Correspondence at: ethomas196@icloud.com}% <-this % stops a space% <-this % stops a space
}

% The paper headers
\markboth{}%
{Shell \MakeLowercase{\textit{et al.}}: Bare Demo of IEEEtran.cls for IEEE Journals}
\maketitle
\begin{abstract}
Cardiovascular diseases, particularly heart failure, are a leading cause of death globally. The early detection of heart failure through routine echocardiogram screenings is often impeded by the high cost and labor-intensive nature of these procedures, a barrier that can mean the difference between life and death. This paper presents ConFormer, a novel, light-weight, convolutional neural network based deep learning model designed to automate the estimation of Left Ventricular Ejection Fraction (LVEF) and Left Ventricular Dimensions (LVD) from echocardiograms.  Results obtained are comparable to SOTA algorithms but with reduced computational complexity.  The implementation of ConFormer has the potential to enhance preventative cardiology by enabling cost-effective, accessible, and comprehensive heart health monitoring.  The source code is available at~\url{https://github.com/Aether111/ConFormer}.
\end{abstract}


\IEEEpeerreviewmaketitle

%#######################################
\section{Introduction}
%#######################################

\IEEEPARstart{T}{he} cardiovascular system, the human circulatory system, consists of various essential organs that circulate oxygen, nutrients, and hormones to all cells and tissues of the body. One of the vital organs in the circulatory system is the heart, which pumps blood throughout the body and receives blood flow back. According to data from the World Health Organization (WHO), cardiovascular disease remains a leading cause of death worldwide. Each year, the death rate from this disease increases, with around 17.9 million people, or 32\% of the world's mortality rate, dying in 2019~\cite{WHO_CVD}. Therefore, a fast and accurate method for cardiac diagnoses is urgently needed for quick and proper management of these conditions.

A common method to assess cardiac function and structure is analysis of echocardiogram videos for estimation of Left Ventricular Ejection Fraction (LVEF) and Left Ventricular Dimensions (LVD).  This assessment serves as the basis for initial screening to diagnose cardiac disease and helps in deciding further treatments.  

However, manually tracing the left ventricle for LVEF and LVD measurements from echocardiogram videos is an expert-dependent tedious task that suffers from inter-observer variability. Moreover, these measurements can vary from one heartbeat to another, and the American Society of Echocardiography (ASE) and the European Association of Cardiovascular Imaging (EACVI) recommend observing up to 5 consecutive heartbeats, making the approach even more complicated.

This paper introduces ConFormer, a novel deep learning model designed to address these challenges.  ConFormer automates the estimation of LVEF and LVD, providing a comprehensive solution for preventative cardiac monitoring that could save countless lives. By automating these critical measurements, ConFormer not only reduces the time and labor required for these assessments, but also eliminates the variability introduced by manual measurements, leading to more accurate and reliable diagnoses.

%#######################################
\section{Previous Work}
%#######################################
The quantification of cardiac function and chamber size is central to cardiac imaging, with echocardiography being the most commonly used noninvasive modality because of its unique ability to provide real-time images of the beating heart, combined with its availability and portability~\cite{Lang2015}.  Moreover, as a cost-effective, radiation-free technique, echocardiography is uniquely suited for deep learning~\cite{Esteva2021}.  While automatic segmentation of the heart region is important to solving practical problems in the field of cardiac medical treatment~\cite{Liu2021}, the majority of deep learning based approaches focus on LV segmentation, with only a few addressing the problem of aortic valve and LA segmentation~\cite{Chen2020}.  In this work, we use deep-learning based segmentation in both of our LVEF and LVD estimation pipelines.

%=====================
\subsection{Datasets}
%=====================
Despite the importance of echocardiography, few broadly validated echocardiograph datasets exist.

The Challenge on Endocardial Three-dimensional Ultrasound Segmentation (CETUS) dataset\footnote{\url{https://www.creatis.insa-lyon.fr/Challenge/CETUS/}} was released in 2014 as part of a competition to compare left ventricle segmentation methods.  The dataset contains 45 3D echocardiograph sequences.  None of the entries implemented deep learning methods.  Next, the Cardiac Acquisitions for Multi-structure Ultrasound Segmentation (CAMUS) dataset\footnote{\url{https://www.creatis.insa-lyon.fr/Challenge/camus/}} was released in 2019 with the goal of providing the largest publicly-available and fully-annotated dataset for 2D echocardiographic assessment using image segmentation and volume estimation.  The dataset contains two and four chamber acquisitions from 500 patients~\cite{Leclerc2019}.  Several of the competition entrants used deep learning based algorithms.  Finally, the Echonet-Dynamic dataset\footnote{\url{https://echonet.github.io/dynamic/}} with 10,030 videos from as many patients, was released in 2019 with labeled measurements of ejection fraction (EF), left ventricular volume at end-systole and end-diastole, and human expert tracings of the left ventricle.  This dataset is claimed by the authors to be the first large release of echocardiogram data with labels and tracings~\cite{Ouyang2019}. 

Besides the EF datasets mentioned above, a comprehensive dataset on LVD measurements is the Echonet-LVH dataset\footnote{\url{https://echonet.github.io/lvh/}}.  This dataset includes 12,000 labeled echocardiogram videos and human expert annotations to provide a baseline to study cardiac chamber size and wall thickness.

In this work, we use both the Echonet-Dynamic and Echonet-LVH datasets for estimation of LVEF and LVD measurements, respectively.  

%=====================
\subsection{LVEF Estimation}
%=====================
LVEF estimation is a cornerstone of modern cardiology~\cite{Marwick2018} with various deep learning based discriminative (convolutional, recurrent, transformer, graph-based) and generative methods reported in the literature.

An initial successful attempt to use deep learning for view classification and cardiac structure segmentation was made by Zhang \textit{et al}~\cite{Zhang2018}.  The authors used a standard VGG-based CNN architecture to classify different echocardiographic views in 14,035 private echocardiograms spanning a 10-year period.  Additionally, a standard U-net based architecture was used to localize cardiac structures.  Ouyang \textit{et al.} used the Echonet-Dynamic dataset to compare various 3D and spatio-temporal factored convolutional architectures for LVEF estimation, and also used a DeepLabV3 architecture for semantic segmentation~\cite{Ouyang2020}.  Taking a different approach, Reynaud \textit{et al.} used a transformer based architecture for LVEF estimation that is not dependent on segmentation~\cite{Reynaud2021}.  

Neural network learning over graph structures has recently drawn a lot of attention.  The first method to use graph neural networks for EF estimation on ultrasound videos used a video encoder, attention encoder and graph regressor~\cite{Mokhtari2022}.  In another first, graph convolutional networks were used for LV segmentation and EF estimation~\cite{Thomas2022}. 

Various works have focused on domain adaptation, an emerging area of importance.  In~\cite{Cai2023}, a generative approach using an image translation and segmentation scheme trained on the CAMUS dataset was tested on the Echonet-Dynamic dataset.  In~\cite{Christensen2023}, a multimodal foundation model using the ConvNeXt image encoder was trained on 1,032,975 unique video-text pairs from 224,685 echocardiography studies for LVEF estimation across 99,870 patients over a decade of clinical care and tested on a different healthcare system.

%=====================
\subsection{LVD Estimation}
%=====================
Besides cardiac function, cardiac geometric measurements also play a central role in cardiac imaging.  The Unity Imaging Collaborative consists of 1894 training images marked with the 4 keypoints required for measuring the LV internal diameter and wall thickness in the PLAX view~\cite{Howard2021}.  The authors trained a feature pyramid based HigherHRNet CNN to detect these 4 keypoints and used it to measure anterior to posterior septum for septal thickness (interventricular septum, IVS), posterior septum to endocardial posterior wall for LV internal diameter (LVID), and endocardial to epicardial posterior wall for posterior wall thickness (LVPW).  They found that the AI was better at matching the expert consensus of the dimension of the LV than it was at choosing the actual keypoint locations.  However, in favor of AI-based methods, the study found that different experts also chose different keypoints for measuring LV dimension.

Duffy \textit{et al.} used human clinician annotations of IVS, LVID and LVPW measurements as training labels to assess ventricular hypertrophy on two transthoracic echocardiography (TTE) views, parasternal long-axis (PLAX) and apical 4-chamber (AP4) 2D videos.  They differentiated hypertrophic cardiomyopathy (HCM) and cardiac amyloid (CA) from other etiologies of increased LV wall thickness~\cite{Duffy2022}.

While most studies use two views, Li \textit{et al.} used six views (AP2, AP3, AP4, PLAX, PSAX-M, PSAX-V) to optimize a pre-trained InceptionResnetV2 model to train a meta-learner under a fusion architecture.  They classified important etyologies (HCM, CA and HTN/others) of increased LV wall thickness~\cite{Li2023}.

% EF architecture
\begin{figure*}[t]
\centering
\begin{subfigure}
[t]{\textwidth}
\centering
\fbox{\includegraphics[width=0.97\textwidth]{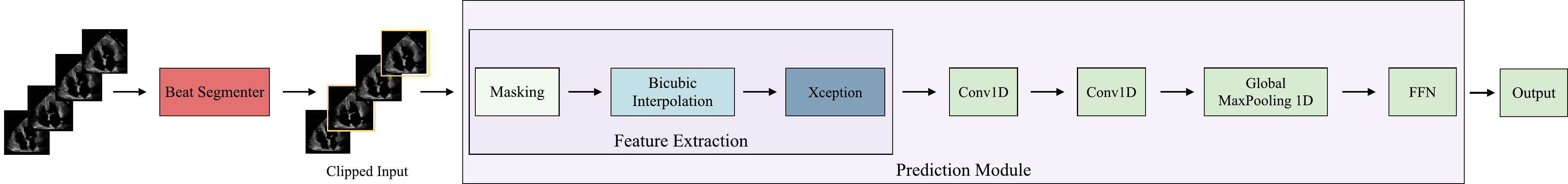}}
%\caption{Architecture for EF estimation.}
\label{fig:EF_arch}
\end{subfigure}

\begin{subfigure}[t]{\textwidth}
\centering
\fbox{\includegraphics[width=0.97\textwidth]{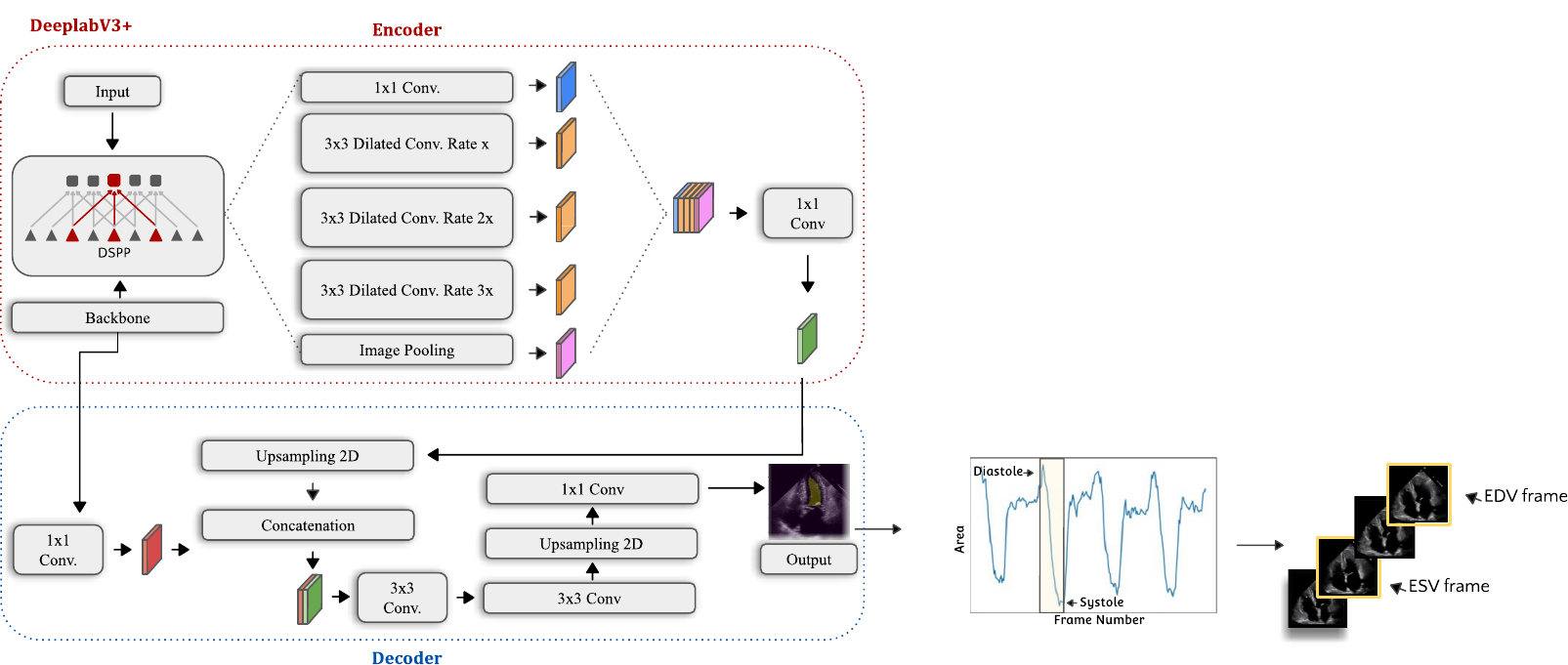}}
%\caption{Beat extractor.}
\label{fig:EF_BS}
\end{subfigure}
\caption{\centering The LVEF prediction pipeline consists of heart-beat extraction, spatial feature extraction, temporal analysis and output regression (top).  The heart-beat extractor is based on the DeepLabV3+ architecture (bottom).}
\label{fig:EF_architecture}
\end{figure*}

%=====================
\subsection{Video Analysis Using CNNs}
%=====================
Several of the methods used in echocardiography video analysis are based on convolutional neural networks, which in turn are widely studied in the context of human action recognition (HAR).  For instance, the influential work on the Echonet-Dynamic dataset by Ouyang \textit{et al.}~\cite{Ouyang2020} is based on the HAR-based R(2+1D) approach of layer-wise spatial and temporal factorization of 3D CNNs~\cite{Tran2018}.   Our work too is based on the related area of network-wise spatio-temporal factorization in FstCN~\cite{Sun2015}.  We therefore study various HAR architectures for our work on video analysis on trans-thoracic echocardiograms.  These architectures fall into two main categories, two stream networks and variants, and 3D convolutions and variants~\cite{Sun2022}. 

In two-stream networks, one stream processes RGB images using 2D convolutions while another stream computes optical flow for motion information~\cite{Simonyan2014}.  Alternately, 3D convolutions are used for jointly modeling spatial and temporal information, such as in~\cite{Ji2012}, C3D~\cite{Tran2015} and I3D~\cite{Carreira2017}.  However, due to the high computational complexity of 3D convolutions, our interest lies in 3D convolutional variants.  

These variants include factoring the 3D convolutions into spatial and temporal convolutions, network-wise, as in FstCN~\cite{Sun2015} on which our work is based, or layer-wise, as in P3D~\cite{Qiu2017}, R(2+1)D~\cite{Tran2018} and Ct-Net~\cite{Li2021}.  For instance, in FstCN~\cite{Sun2015}, network factorization is used with the spatial convolutional layers followed by two temporal convolutional layers.  Layer factorization is used in R(2+1)D~\cite{Tran2018}, and each spatio-temporal convolution is factored into a block of spatial and temporal convolution.  This architecture therefore alternates between spatial and temporal convolutions.  In Ct-Net~\cite{Li2021}, the spatial, temporal and channel dimensions are treated separately, making C3D and R(2+1)D special cases of this architecture. Other 3D convolutional variants include using both 3D and 2D convolutions together, as in S3D~\cite{Xie2018}, or using temporal modules with 2D CNNs, as in Tsm, Tdn and Action-net~\cite{Lin2019, Wang2021L, Wang2021Z}. 

We now explain the details of both our LVEF and LVD methods.

%#######################################
\section{LVEF (Left Ventricular Ejection Fraction) Estimation Model}
%#######################################
Ejection fraction (EF) is a key indicator of both heart function and structural changes in the heart, and is broadly acknowledged as an essential tool for diagnosis and prognosis. Its application across diverse clinical scenarios, including heart failure, myocardial infarction, and valvular heart diseases, has established it as a fundamental element in contemporary cardiology, influencing both guidelines and clinical practice~\cite{Marwick2018}.  

In order to compute the EF in standard echocardiography practice, the left ventricle is traced along the endocardial border at end-systole and end-diastole. These areas are then integrated across the length of the ventricle's major axis to compute the left ventricular end-systolic volume (ESV) and the left ventricular end-diastolic volume (EDV), from which EF is computed as~\cite{Ouyang2019}:

\begin{equation}
EF (\%) = \frac{EDV - ESV}{EDV}
\end{equation}

%====================
\subsection{Data}
%====================
The data used for this study was the EchoNet-Dynamic dataset, published by Stanford University~\cite{Ouyang2019}. The dataset contains 10,030 labeled apical 4-chamber (AP4) echocardiogram videos from unique patients at Stanford University Medical Center. The videos cover a variety of typical lab imaging acquisition conditions and include human expert annotations such as LVEF measurements and tracings of the left ventricle.

%====================
\subsection{Design}
%====================
The ConFormer model was designed to leverage the strengths of Convolutional Neural Networks (CNNs). The model is fully automatic, accepting raw echocardiogram videos without the need for manual pre-processing and consists of a Beat Extractor and a Prediction Module.  The architecture is based on a flexible, decoupled spatial, channel and temporal convolutional design.  The spatial and channel decoupling is achieved through the use of the depth-wise separable convolutions of the Xception backbone~\cite{Chollet2017}.  Even though this backbone is pre-trained on the ImageNet dataset, our design nevertheless inherits the flexibility of independently modifying the spatial and channel convolutions if required.

%----------------
\subsubsection{Decoupling the Cross-Channel Convolutions}
%----------------
The Inception architecture family built on previous work and explicitly factored spatial and cross-channel correlations~\cite{Szegedy2015}.  This family of models first demonstrated the advantages of factoring convolutions into multiple branches operating successively on channels and then on space.  The next step in this evolution was ``extreme'' Inception, i.e., Xception, that showed how to scale-up depth-wise separable convolutions in convolutional neural networks~\cite{Chollet2017}.  At approximately 23 million parameters each, Xception out-performed Inception V3 on the ImageNet and JFT datasets.  Since depth-wise separable convolutions are a drop-in replacement for standard convolutions that empirically work almost as well but with reduced computational complexity, they were adopted by the more frugal 4.2 million parameter MobileNet architecture~\cite{Howard2017}. However, we have chosen to use the Xception backbone, a linear stack of depth-wise separable convolutional layers with residual connections, as a happy compromise between complexity and efficiency in our decoupled architecture.  

% EF segmentation
\begin{figure}[t]
\newcommand{\figw}{0.1}
\centering
\begin{subfigure}[t]{\figw\textwidth}
\includegraphics[width=\textwidth]{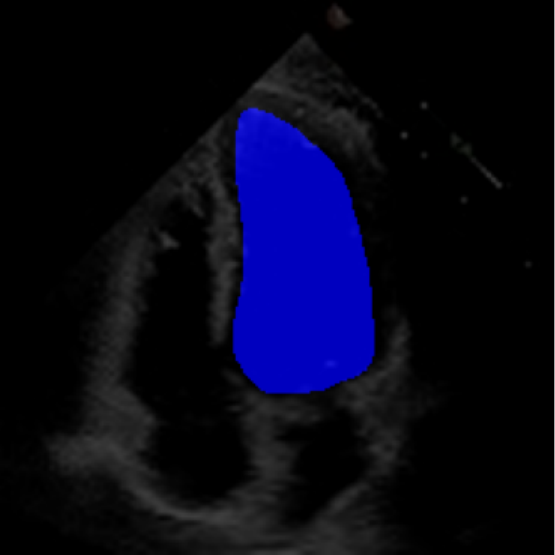}
\label{fig:seg_fig0}
\end{subfigure}\begin{subfigure}[t]{\figw\textwidth}
\includegraphics[width=\textwidth]{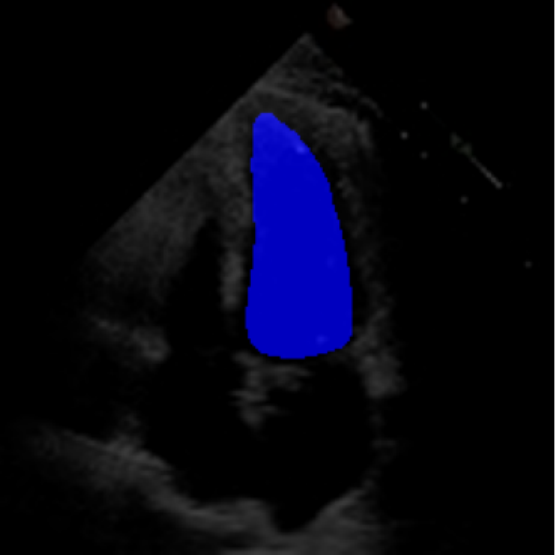}
\label{fig:seg_fig1}
\end{subfigure}
\begin{subfigure}[t]{\figw\textwidth}
\includegraphics[width=\textwidth]{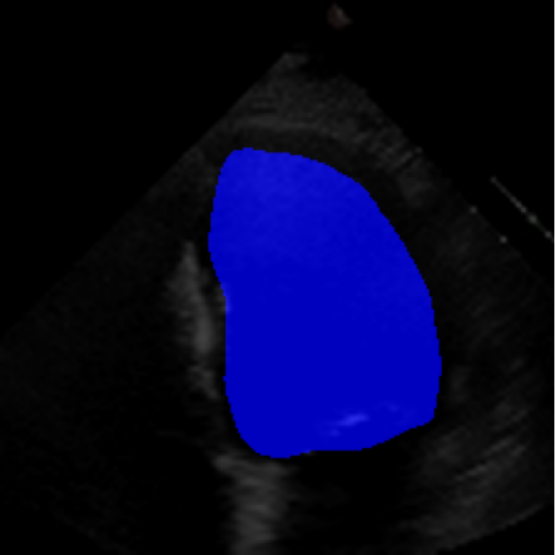}
\label{fig:seg_fig2}
\end{subfigure}\begin{subfigure}[t]{\figw\textwidth}
\includegraphics[width=\textwidth]{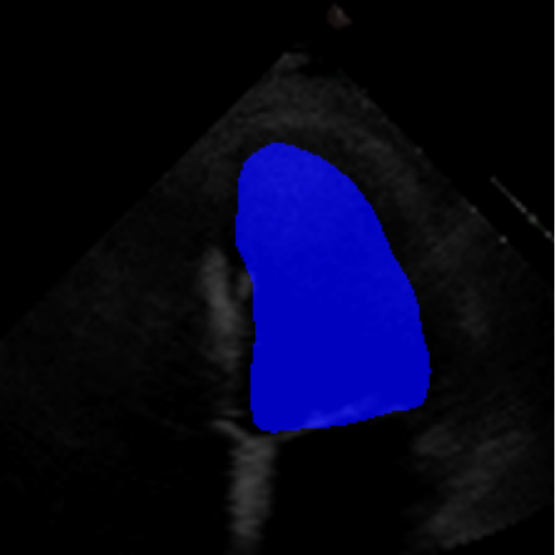}
\label{fig:seg_fig3}
\end{subfigure}
\begin{subfigure}[t]{\figw\textwidth}
\includegraphics[width=\textwidth]{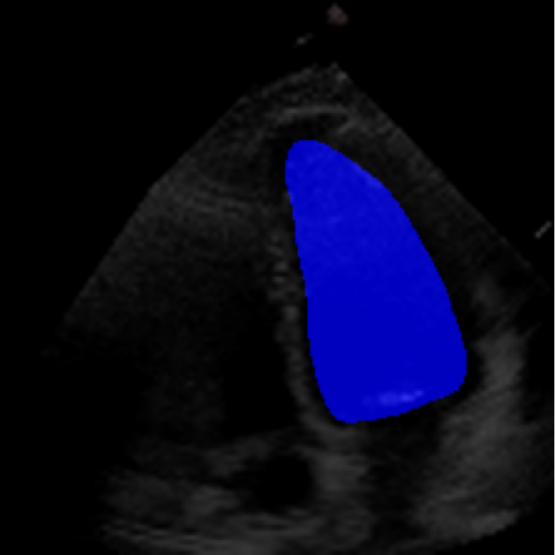}
\label{fig:seg_fig4}
\end{subfigure}\begin{subfigure}[t]{\figw\textwidth}
\includegraphics[width=\textwidth]{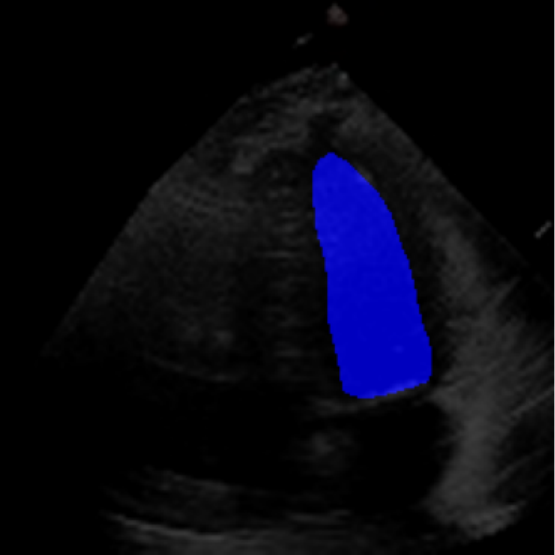}
\label{fig:seg_fig5}
\end{subfigure}
\begin{subfigure}[t]{\figw\textwidth}
\includegraphics[width=\textwidth]{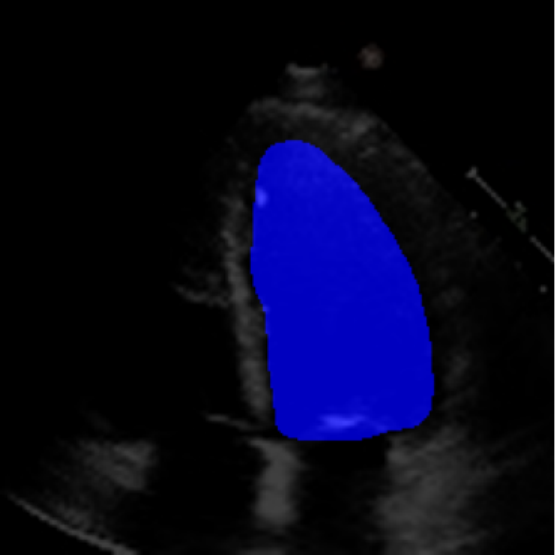}
\label{fig:seg_fig8}
\end{subfigure}\begin{subfigure}[t]{\figw\textwidth}
\includegraphics[width=\textwidth]{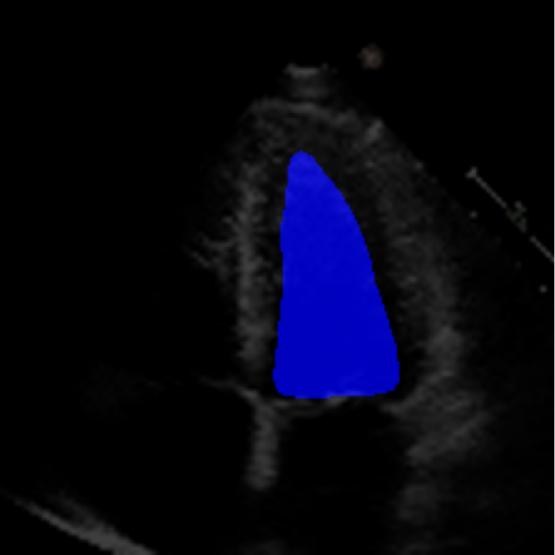}
\label{fig:seg_fig9}
\end{subfigure}

\begin{subfigure}[t]{0.48\textwidth}
\includegraphics[width=\textwidth]{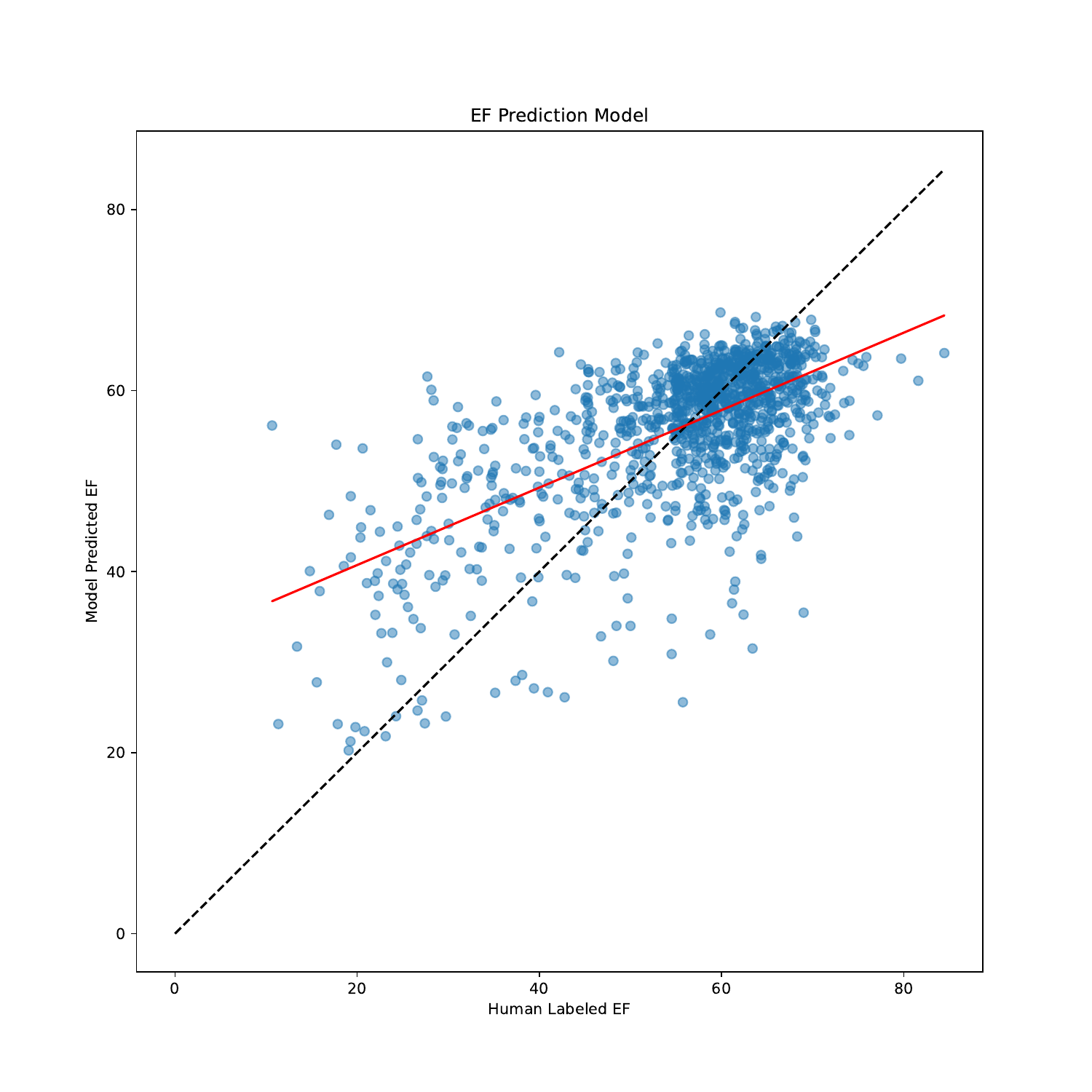}
\end{subfigure}
\caption{Segmentation of left ventricular cardiac chamber during diastole (left figures) and systole (rights figures) during the same heartbeat (above).  Predicted vs manually calculated ejection fraction (below).}
\label{fig:EF_pred_vs_manual}
\end{figure}

%----------------
\subsubsection{Decoupling the Temporal Convolutions}
%----------------
In 3D convolution on spatial and temporal domains, a video signal $\mathbf{V}~\in~\mathbb{R}^{n_x\times {n_y} \times n_t}$ is convolved with a 3D kernel $\mathbf{K}\in \mathbb{R}^{m_x\times {m_y} \times m_t}$ to get features $\mathbf{F}$,

\begin{equation}
\mathbf{F} =  \mathbf{V} \ast \mathbf{K}
\end{equation}

This is a computationally expensive operation, and can be factored, under certain conditions, into a spatial convolution followed by a temporal convolution.  Specifically, this can be done if the kernel $\mathbf{K}$ can be factorized into a Kronecker product of a 2D spatial kernel $\mathbf{K}_s$ and a 1D temporal kernel $\mathbf{k}_t$.  

\begin{equation}
\mathbf{K} = \mathbf{K_s} \otimes \mathbf{k_t}
\end{equation}

Besides reducing computational complexity, an added benefit of this factorization is that if the 2D spatial kernel $\mathbf{K}_s$ is pre-trained, then we only need to train the temporal kernel $\mathbf{k}_t$.  Consequently, the 3D convolution can then be carried out in 2 steps.  At time $t$, the spatial convolution is given by

\begin{equation}
\mathbf{F}_s(:,:,t) = \mathbf{V}(:,:,t) \ast \mathbf{K}_s
\end{equation}

This is followed by temporal convolution,

\begin{equation}
\mathbf{F}_{st}(x,y,:) =  \mathbf{F}_s(x,y,:) \ast \mathbf{k}_t
\end{equation}

Here, $\mathbf{V}(:,:,t)$ is a single frame.  Each individual dot product in the 3D convolution has computational complexity of  $m_x \times m_y \times m_t$.  In the factored convolution, this computational complexity reduces to $(m_x \times m_y) + m_t$.  Since convolution is essentially a sliding dot product, this process is repeated $n_x \times n_y$ times leading to substantial computational savings.  As mentioned earlier, further savings come from using pre-trained 2D spatial kernels.

\begin{table}[t]
\centering
\begin{tabular}{|l|l|l|l|l|}\hline
\textbf{Model}  & \textbf{Evaluation} & \textbf{Parameters} & \textbf{MAE}\\\hline 
Echonet-Dynamic &  Beat-by-beat & 71.1 mil & 4.05\%\\\hline
Echonet-Dynamic &  All frames & - & 7.35\%\\\hline
UVT-M &  Beat-by-beat  & 346.8 mil & 5.32\%\\\hline
UVT-R &  128 frames  & 346.8 mil & 5.95\%\\\hline
UVT-R &  128 frames  & 346.8 mil & 6.77\%\\\hline
\textbf{Conformer} & \textbf{All frames} & \textbf{5.82 mil} & \textbf{6.57\%} \\\hline
\end{tabular}
\caption{Results}
\label{table:results}
\end{table}
%----------------
\subsubsection{Beat Extractor}
%----------------
The 112x112 input is first resized to 299x299 as required by the Xception model.  The Beat Extractor is then used to feed only whole beats to the LVEF estimation pipeline.  A modified DeepLabV3+ model segments the left ventricle for every frame of the echocardiogram video (Figure~\ref{fig:segmentation}).  The area is measured, and a peak detector is then used to find the maximum (diastole) and minimum area (systole).  The frame numbers corresponding to these extrema are recorded, and the video is clipped into smaller videos that go from diastole to systole. These smaller videos are independently fed to the Xception based feature extractor.

%\begin{equation}
%    \mathbf{y}[i] = \sum_k x[i + r \cdot k]w[k]
%\end{equation}

\begin{figure*}[t]
\centering
\fbox{\includegraphics[width=0.98\textwidth]{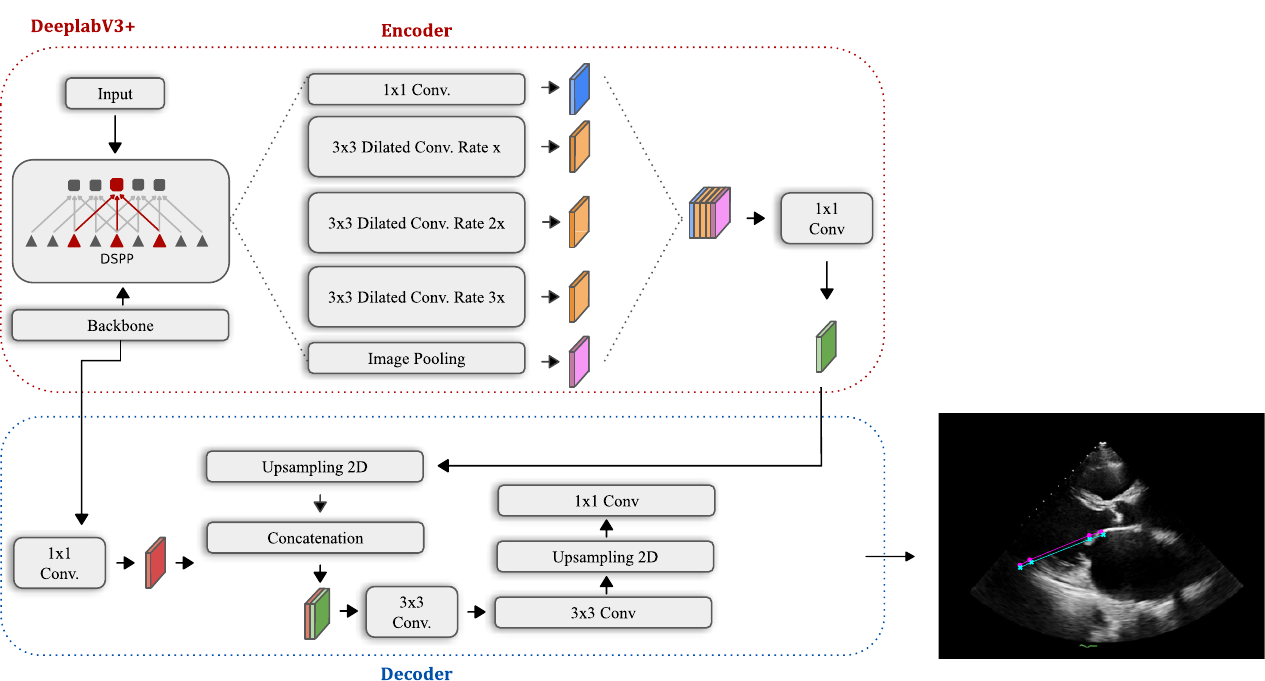}}
\caption{The LVD (left ventricular dimensions) prediction pipeline.}
\label{fig:arch_LVH}
\end{figure*}

%----------------
\subsubsection{Prediction Module}
%----------------
The Prediction Module processes the shortened clips of a beat (diastole to systole).  This module includes an Xception feature extractor (for processing spatial features), a 1D convolutional stack composed of two 1D convolutional layers (for learning temporal dependencies), a feed forward Network (FFN) composed of two dense layers, and a regressive output neuron that estimates EF. The entire model is highly efficient, with just 5.82 million parameters in total.  The performance was evaluated based on the mean absolute error of the predicted Ejection Fraction (EF).

The batch size is set at 64.  The output of the Xception module is a 64$\times$2048 dimensional vector.  This is passed through a 128-filter 1D convolutional layer with a kernel size of 7 elements.  The average frame rate of the input videos is 51.1 frames per second with 90\% of the videos lying within 0.95 standard deviations of this average frame rate.  At 51 frames per second, the average heart beat (0.8 sec) spans about 41 frames. Our temporal convolutional length of 7 therefore processes about 1/6 th of a heart beat at a time.  This is sufficiently large to capture temporal variations in the heart beat cycle.  The second 1D convolutional layer has 256 filters with kernel size of 5.  The results of these 1D filters are then passed through a global max pool, two layers of 256 fully connected neurons with Swish activation, and onto a regression layer to finally predict the LVEF.  Hyperparameter optimization was carried out to determine convolutional layer and kernel sizes, as well as dense layer sizes and activations.  Training on an Nvidia A-100 GPU took under one hour.

%====================
\subsection{Results and Discussion}
%====================
The ConFormer model demonstrated high performance in the estimation of Left Ventricular Ejection Fraction (LVEF) from echocardiogram videos (Figure~\ref{fig:EF_pred_vs_manual}). The model achieved a mean absolute error of 6.57 in LVEF prediction, indicating a high level of accuracy in its estimations (Table~\ref{table:results}). Notably, ConFormer outperforms the full video assessment of LVEF by EchoNet-Dynamic, a significant achievement given the model's efficiency. With just 5.82 million parameters, ConFormer is significantly more efficient than existing models, making it a practical tool for real-world applications.

While the estimation of LVEF is a critical component of heart health monitoring, a comprehensive solution also requires the measurement of Left Ventricular Dimensions. Wall thickness measurements provide additional information about the heart's structure and function, complementing the LVEF estimation to provide a more complete picture of heart health.

The results of this study suggest that ConFormer could enhance preventative cardiology by enabling cost-effective, accessible, and comprehensive heart health monitoring. By automating the estimation of LVEF and LVD, ConFormer not only reduces the time and labor required for these assessments but also eliminates the variability introduced by manual measurements, leading to more accurate and reliable diagnoses.

However, it's important to note that while the results are promising, further research is needed to validate the model's performance in different clinical settings and populations. Future work could also explore the integration of ConFormer with other diagnostic tools to further enhance its utility in heart health monitoring.

% LVH results
\begin{figure}[t]
\newcommand{\figw}{0.2}
\centering
\begin{subfigure}[t]{\figw\textwidth}
\includegraphics[width=\textwidth]{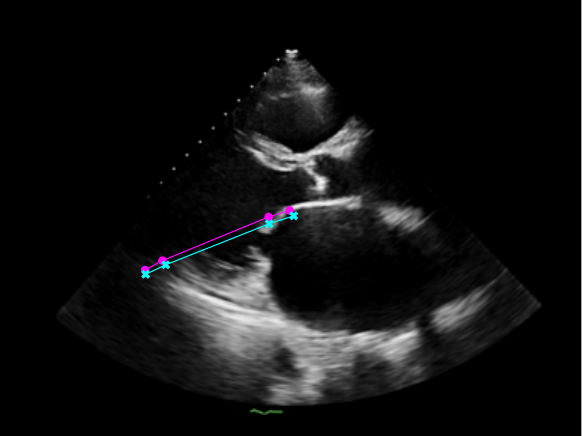}
\label{fig:LVH_20}
\end{subfigure} \begin{subfigure}[t]{\figw\textwidth}
\includegraphics[width=\textwidth]{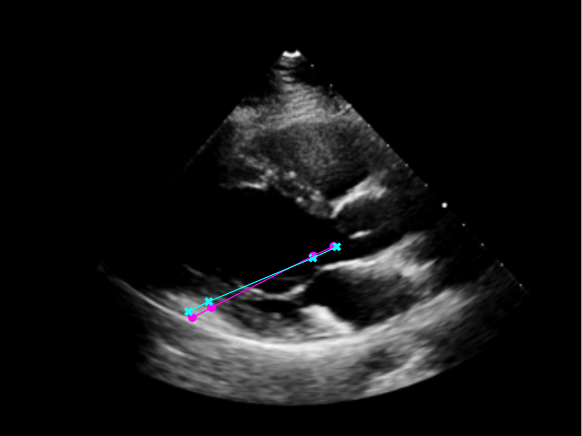}
\label{fig:LVH_126}
\end{subfigure}
\begin{subfigure}[t]{\figw\textwidth}
\includegraphics[width=\textwidth]{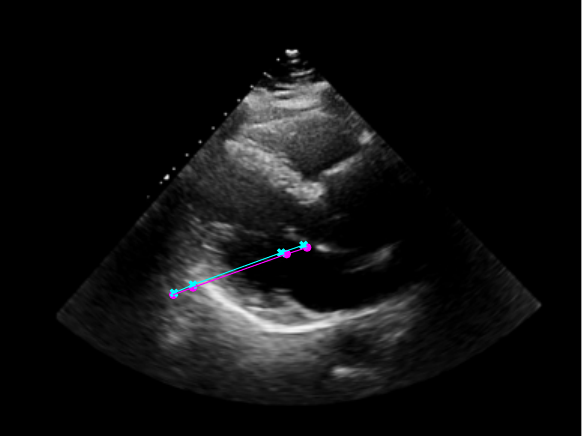}
\label{fig:LVH_165}
\end{subfigure} \begin{subfigure}[t]{\figw\textwidth}
\includegraphics[width=\textwidth]{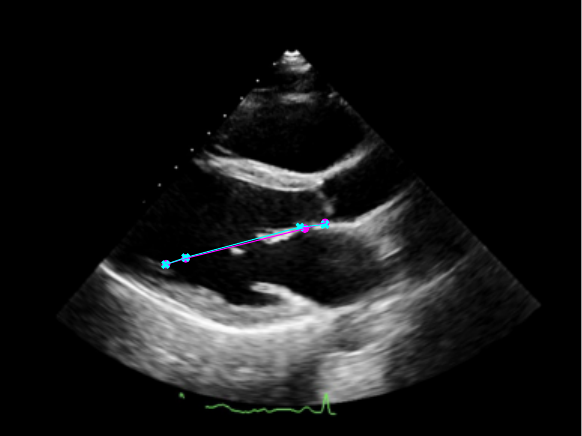}
\label{fig:LVH_200}
\end{subfigure}
\caption{LV dimensions.}
\label{fig:LV_dimensions}
\end{figure}

%#######################################
\section{LVD (Left Ventricular Dimensions) Estimation Model}
%#######################################
%====================
\subsection{Data}
%====================
The data used for this study was the EchoNet-LVH dataset, published by Stanford University~\cite{Duffy2022}.  The dataset contains 12,001 parasternal long-axis (PLAX) echocardiogram videos with human clinician annotations of intraventricular septum (IVS), LV internal dimension (LVID) and LV posterior wall (LVPW) measurements.  These dimensions can be used to quantify ventricular wall thickness, and predict the cause of left ventricular hypertropy (LVH).

%====================
\subsection{Design}
%====================
The model for LVD measurement was built using a modified version of the DeepLabV3+ architecture~\cite{Chen2018}, with an EfficientNetV2S~\cite{Tan2021} backbone replacing the original ResNet50~\cite{He2016}. This change was necessitated by the higher resolution of the EchoNet-LVH videos, which typically have dimensions of either (1024,768) or (800,600), compared to the (112,112) resolution of the Echonet-Dynamic dataset.

As seen in Figure~\ref{fig:LV_dimensions}, the model was trained to identify four key points in the echocardiogram videos: the beginning and end of the intraventricular septum (IVS), and the beginning and end of the left ventricular posterior wall (LVPW). These points were used to calculate the lengths of the IVS and LVPW, and the internal diameter of the left ventricle (LVID).

The model was compiled using a custom loss function, which was a weighted combination of the mean squared errors of the predicted and actual lengths of the IVS, LVPW, and LVID. The weights were determined by the inverse of the standard deviations of the lengths in the training set, to give more importance to the measurements that had less variability.

The model for wall thickness measurement was built using a modified version of the DeepLabV3+ architecture, with an EfficientNetV2S backbone replacing the original ResNet50. This change was necessitated by the higher resolution of the EchoNet-LVH videos, which typically have dimensions of either (1024,768) or (800,600), compared to the (112,112) resolution of the Echonet-Dynamic dataset.

As seen in Figure~\ref{fig:LV_dimensions}, the model was trained to identify four key points in the echocardiogram videos: the beginning and end of the intraventricular septum (IVS), and the beginning and end of the left ventricular posterior wall (LVPW). These points were used to calculate the lengths of the IVS and LVPW, and the internal diameter of the left ventricle (LVID).

%====================
\subsection{Results and Discussion}
%====================
The ConFormer model achieved a mean absolute error (MAE) of approximately 6 for the combined measurements of the IVS, LVID, and LVPW. This performance is notable when compared to the EchoNet-LVH model, which achieved MAEs of 1.7mm for IVS, 3.8mm for LVID, and 1.8mm for LVPW when fine-tuned on the Cedars-Sinai Medical Center (CSMC) dataset.

Despite its significantly reduced parameter count, the ConFormer model demonstrates robust performance across different video resolutions and frame rates. These results suggest that the ConFormer model could be a valuable tool for automating the measurement of ventricular wall thickness in echocardiogram videos, potentially leading to more accurate and efficient diagnoses of left ventricular hypertrophy.

%#######################################
\section{Conclusion}
%#######################################
In conclusion, the ConFormer model demonstrates the potential of compact and efficient deep learning models for automated measurement of ejection fraction and wall thickness in echocardiograms. Despite its significantly reduced parameter count, ConFormer achieves comparable performance to existing models, making it a practical tool for real-world applications. The model's robust performance across different video resolutions and frame rates further underscores its versatility and potential for use in a variety of clinical settings.

The results of this study suggest that automated analysis of echocardiograms using deep learning models like ConFormer could lead to more accurate and efficient diagnoses of cardiovascular diseases, potentially saving lives by enabling earlier detection and treatment. Future work could explore the application of ConFormer to other tasks in echocardiogram analysis, as well as its integration into clinical workflows.

%#######################################
\section*{Acknowledgment}
%#######################################
The authors would like to thank Eris Thomas for being a crucial part of the research process.

\ifCLASSOPTIONcaptionsoff
  \newpage
\fi

\printbibliography
\end{document}